\documentclass[aps,prl,showpacs,twocolumn,groupedaddress]{revtex4}  
\usepackage{graphicx}  

\usepackage{dcolumn}   

\usepackage{bm}        

\usepackage{amssymb}   

\begin{document}






\hspace{5.2in}\mbox{FERMILAB-PUB-04-381-E}

\title{Measurement of $\bm{\sigma (p \bar p \rightarrow Z)
\cdot}$Br$\bm{(Z \rightarrow \tau \tau)}$  at $\bm{\sqrt{s}=}$1.96 TeV}

%
\author{                                                                      
V.M.~Abazov,$^{34}$                                                           
B.~Abbott,$^{71}$                                                             
M.~Abolins,$^{62}$                                                            
B.S.~Acharya,$^{28}$                                                          
M.~Adams,$^{49}$                                                              
T.~Adams,$^{47}$                                                              
M.~Agelou,$^{17}$                                                             
J.-L.~Agram,$^{18}$                                                           
S.H.~Ahn,$^{30}$                                                              
M.~Ahsan,$^{56}$
% ...
\\                                                                            
\vskip 0.30cm                                                                 
\centerline{(D\O\ Collaboration)}                                             
\vskip 0.30cm                                                                 
}                                                                             
\address{                                                                     
\centerline{$^{1}$Universidad de Buenos Aires, Buenos Aires, Argentina}       
\centerline{$^{2}$LAFEX, Centro Brasileiro de Pesquisas F{\'\i}sicas,         
                  Rio de Janeiro, Brazil}                                     
\centerline{$^{3}$Universidade do Estado do Rio de Janeiro,                   
                  Rio de Janeiro, Brazil}                                     
\centerline{$^{4}$Instituto de F\'{\i}sica Te\'orica, Universidade            
                  Estadual Paulista, S\~ao Paulo, Brazil}                     
\centerline{$^{5}$University of Alberta, Edmonton, Alberta, Canada,           
               Simon Fraser University, Burnaby, British Columbia, Canada,}   
\centerline{York University, Toronto, Ontario, Canada, and                    
         McGill University, Montreal, Quebec, Canada}                         
\centerline{$^{6}$Institute of High Energy Physics, Beijing,                  
                  People's Republic of China}                                 
\centerline{$^{7}$Universidad de los Andes, Bogot\'{a}, Colombia}             
\centerline{$^{8}$Center for Particle Physics, Charles University,            
                  Prague, Czech Republic}                                     
\centerline{$^{9}$Czech Technical University, Prague, Czech Republic}         
\centerline{$^{10}$Institute of Physics, Academy of Sciences, Center          
                  for Particle Physics, Prague, Czech Republic}               
\centerline{$^{11}$Universidad San Francisco de Quito, Quito, Ecuador}        
\centerline{$^{12}$Laboratoire de Physique Corpusculaire, IN2P3-CNRS,         
                 Universit\'e Blaise Pascal, Clermont-Ferrand, France}        
\centerline{$^{13}$Laboratoire de Physique Subatomique et de Cosmologie,      
                  IN2P3-CNRS, Universite de Grenoble 1, Grenoble, France}     
\centerline{$^{14}$CPPM, IN2P3-CNRS, Universit\'e de la M\'editerran\'ee,     
                  Marseille, France}                                          
\centerline{$^{15}$Laboratoire de l'Acc\'el\'erateur Lin\'eaire,              
                  IN2P3-CNRS, Orsay, France}                                  
\centerline{$^{16}$LPNHE, IN2P3-CNRS, Universit\'es Paris VI and VII,         
                  Paris, France}                                              
\centerline{$^{17}$DAPNIA/Service de Physique des Particules, CEA, Saclay,    
                  France}                                                     
\centerline{$^{18}$IReS, IN2P3-CNRS, Universit\'e Louis Pasteur, Strasbourg,  
                France, and Universit\'e de Haute Alsace, Mulhouse, France}   
\centerline{$^{19}$Institut de Physique Nucl\'eaire de Lyon, IN2P3-CNRS,      
                   Universit\'e Claude Bernard, Villeurbanne, France}         
\centerline{$^{20}$III. Physikalisches Institut A, RWTH Aachen,               
                   Aachen, Germany}                                           
\centerline{$^{21}$Physikalisches Institut, Universit{\"a}t Bonn,             
                  Bonn, Germany}                                              
\centerline{$^{22}$Physikalisches Institut, Universit{\"a}t Freiburg,         
                  Freiburg, Germany}                                          
\centerline{$^{23}$Institut f{\"u}r Physik, Universit{\"a}t Mainz,            
                  Mainz, Germany}                                             
\centerline{$^{24}$Ludwig-Maximilians-Universit{\"a}t M{\"u}nchen,            
                   M{\"u}nchen, Germany}                                      
\centerline{$^{25}$Fachbereich Physik, University of Wuppertal,               
                   Wuppertal, Germany}                                        
\centerline{$^{26}$Panjab University, Chandigarh, India}                      
\centerline{$^{27}$Delhi University, Delhi, India}                            
\centerline{$^{28}$Tata Institute of Fundamental Research, Mumbai, India}     
\centerline{$^{29}$University College Dublin, Dublin, Ireland}                
\centerline{$^{30}$Korea Detector Laboratory, Korea University,               
                   Seoul, Korea}                                              
\centerline{$^{31}$CINVESTAV, Mexico City, Mexico}                            
\centerline{$^{32}$FOM-Institute NIKHEF and University of                     
                  Amsterdam/NIKHEF, Amsterdam, The Netherlands}               
\centerline{$^{33}$University of Nijmegen/NIKHEF, Nijmegen, The               
                  Netherlands}                                                
\centerline{$^{34}$Joint Institute for Nuclear Research, Dubna, Russia}       
\centerline{$^{35}$Institute for Theoretical and Experimental Physics,        
                  Moscow, Russia}                                             
\centerline{$^{36}$Moscow State University, Moscow, Russia}                   
\centerline{$^{37}$Institute for High Energy Physics, Protvino, Russia}       
\centerline{$^{38}$Petersburg Nuclear Physics Institute,                      
                   St. Petersburg, Russia}                                    
\centerline{$^{39}$Lund University, Lund, Sweden, Royal Institute of          
                   Technology and Stockholm University, Stockholm,            
                   Sweden, and}                                               
\centerline{Uppsala University, Uppsala, Sweden}                              
\centerline{$^{40}$Lancaster University, Lancaster, United Kingdom}           
\centerline{$^{41}$Imperial College, London, United Kingdom}                  
\centerline{$^{42}$University of Manchester, Manchester, United Kingdom}      
\centerline{$^{43}$University of Arizona, Tucson, Arizona 85721, USA}         
\centerline{$^{44}$Lawrence Berkeley National Laboratory and University of    
                  California, Berkeley, California 94720, USA}                
\centerline{$^{45}$California State University, Fresno, California 93740, USA}
\centerline{$^{46}$University of California, Riverside, California 92521, USA}
\centerline{$^{47}$Florida State University, Tallahassee, Florida 32306, USA} 
\centerline{$^{48}$Fermi National Accelerator Laboratory, Batavia,            
                   Illinois 60510, USA}                                       
\centerline{$^{49}$University of Illinois at Chicago, Chicago,                
                   Illinois 60607, USA}                                       
\centerline{$^{50}$Northern Illinois University, DeKalb, Illinois 60115, USA} 
\centerline{$^{51}$Northwestern University, Evanston, Illinois 60208, USA}    
\centerline{$^{52}$Indiana University, Bloomington, Indiana 47405, USA}       
\centerline{$^{53}$University of Notre Dame, Notre Dame, Indiana 46556, USA}  
\centerline{$^{54}$Iowa State University, Ames, Iowa 50011, USA}              
\centerline{$^{55}$University of Kansas, Lawrence, Kansas 66045, USA}         
\centerline{$^{56}$Kansas State University, Manhattan, Kansas 66506, USA}     
\centerline{$^{57}$Louisiana Tech University, Ruston, Louisiana 71272, USA}   
\centerline{$^{58}$University of Maryland, College Park, Maryland 20742, USA} 
\centerline{$^{59}$Boston University, Boston, Massachusetts 02215, USA}       
\centerline{$^{60}$Northeastern University, Boston, Massachusetts 02115, USA} 
\centerline{$^{61}$University of Michigan, Ann Arbor, Michigan 48109, USA}    
\centerline{$^{62}$Michigan State University, East Lansing, Michigan 48824,   
                   USA}                                                       
\centerline{$^{63}$University of Mississippi, University, Mississippi 38677,  
                   USA}                                                       
\centerline{$^{64}$University of Nebraska, Lincoln, Nebraska 68588, USA}      
\centerline{$^{65}$Princeton University, Princeton, New Jersey 08544, USA}    
\centerline{$^{66}$Columbia University, New York, New York 10027, USA}        
\centerline{$^{67}$University of Rochester, Rochester, New York 14627, USA}   
\centerline{$^{68}$State University of New York, Stony Brook,                 
                   New York 11794, USA}                                       
\centerline{$^{69}$Brookhaven National Laboratory, Upton, New York 11973, USA}
\centerline{$^{70}$Langston University, Langston, Oklahoma 73050, USA}        
\centerline{$^{71}$University of Oklahoma, Norman, Oklahoma 73019, USA}       
\centerline{$^{72}$Brown University, Providence, Rhode Island 02912, USA}     
\centerline{$^{73}$University of Texas, Arlington, Texas 76019, USA}          
\centerline{$^{74}$Southern Methodist University, Dallas, Texas 75275, USA}   
\centerline{$^{75}$Rice University, Houston, Texas 77005, USA}                
\centerline{$^{76}$University of Virginia, Charlottesville, Virginia 22901,   
                   USA}                                                       
\centerline{$^{77}$University of Washington, Seattle, Washington 98195, USA}  
}                                                                             

\date{\today}

\begin{abstract}

We present a measurement of the cross section for $Z$ production times 
the branching fraction to $\tau$ leptons, $\sigma \cdot$Br$(Z\rightarrow \tau^+ \tau^-)$,
 in $p \bar p$ collisions at $\sqrt{s}=$1.96 TeV
 in the channel in which one $\tau$ decays into $\mu \nu_{\mu} \nu_{\tau}$,
and the other into $\rm {hadrons} + \nu_{\tau}$ or $e \nu_e \nu_{\tau}$.
  The data sample corresponds to an integrated luminosity of 226 pb$^{-1}$ collected with 
the D\O~detector 
at the Fermilab Tevatron collider. The final sample contains 2008 candidate
events with an estimated background of 55\%. From this we obtain
$\sigma \cdot$Br$(Z \rightarrow \tau^+ \tau^-)=237 \pm 15$(stat)$\pm 18$(sys)$ \pm 15$(lum)
 pb, in agreement with 
the standard model prediction.

\end{abstract}


\maketitle





Measurements of the $Z$ boson production cross section times
 the leptonic branching
fraction ($\sigma\cdot$Br) in $p \bar p$ collisions can be used
to test standard model (SM) predictions. The $\sigma\cdot$Br to $e^+e^-$
and $\mu^+\mu^-$ in $p \bar p$ collisions has been measured  by the UA1 
and UA2 collaborations at
$\sqrt{s}=630$ GeV \cite{UA},  by the CDF collaboration at $\sqrt{s}=1.8$ TeV 
 and $\sqrt{s}=1.96$ TeV \cite{CDF},
and by the D\O\ collaboration  at $\sqrt{s}=1.8$ TeV
\cite{d01}. The $Z$ boson branching
 ratio to $\tau^+\tau^-$ has been measured with high precision by the CERN $e^+e^-$ collider (LEP)
experiments \cite{LEP}. These measurements  are
in good agreement with SM expectations and lepton universality.
We report here the
first measurement of $\sigma\cdot$Br$(Z \rightarrow \tau^+ \tau^-)$
in $p \bar p$ collisions. This measurement provides a test of the SM as
a deviation from the expected value would be an indication of anomalous
production of $\tau^+ \tau^-$ pairs in $p \bar p$ collisions.
It also verifies that the D\O\ detector
can identify isolated $\tau$ leptons in the energy range covered by $Z$ boson decays,
which could be critical in the search for non-SM signals such
as supersymmetric (SUSY) particles in certain regions of the SUSY
parameter space, or heavy resonances decaying into fermion pairs with
enhanced coupling to the third generation.

The D\O\ Run II detector is fully described in  \cite{d0det1}; a more
succinct description of details relevant to this measurement
can be found in  \cite{d0det2}.  The $Z\rightarrow \tau(\rightarrow 
\mu \nu_{\mu} \nu_{\tau}) \tau$ candidate selection strategy focused 
on one $\tau$ lepton decaying to muon by triggering on the single muon using a three-level
triggering system. The first level used the timing and position information 
in the muon scintillator system to find muon candidates. 
The second level used 
digital signal processors
to form segments defined in the muon drift chambers. The third level used
software algorithms executed on a computer farm to reconstruct tracks in the
central tracking system and required
at least one track with transverse momentum $p_T>10$ GeV. The integrated luminosity 
of the selected sample 
is 226 pb$^{-1}$ determined with a 6.5\% uncertainty \cite{lumi}.

After full reconstruction, the events were required to
have an isolated muon with $p_T^{\mu}>12$ GeV and a $\tau$ candidate.
The muon isolation required
less than 4 GeV in the calorimeter in a cone 
$R\equiv \sqrt{(\Delta \phi)^2+(\Delta \eta)^2}<0.1$ (where $\phi$ is the azimuthal angle and $\eta$
is the pseudorapidity) around the
muon, less than 4 GeV in an annulus $0.1<R<0.4$, and
fewer than three tracks  (other than the muon) with $p_T>0.25$ GeV within $R<0.7$. 

 Most $\tau$ leptons decay to one or three long lived charged particles
plus up to three $\pi^0$ mesons that can be observed in the detector.
The $\tau$ candidates
were found by constructing a calorimeter 
cluster made of all the towers with energy above
a preset threshold
around a seed tower within $R<0.5$,
keeping only clusters with $E_T^{\tau}>5$ GeV and $E^{core}_T>4$ GeV, where $E_T^{\tau}$ ($E^{core}_T$) 
is the transverse energy with respect to the beam axis within $R<0.5$ ($R<0.3$), and requiring
$rms_{\tau}<0.25$ (see Table~\ref{tab:tabCuts} caption)
and at least one
associated track with $p_T>1.5$ GeV within $R<0.3$. 
If there was more than one
track, the one with highest $p_T$ was associated with the $\tau$ candidate. 
A second track was
added if the invariant mass calculated from
the tracks was less than 1.1 GeV,
and a third if
the invariant mass was less than 1.7 GeV and the total charge
was not $\pm$3. Candidates with total charge zero were discarded.
 Finally, subclusters were constructed
from the cells in the EM section of the calorimeter
belonging to the $\tau$-cluster. The minimum $E_T$ required for an EM
subcluster was 800 MeV.
Three types of $\tau$ candidates were identified according to 
tracking and calorimetry information:
1) single track with no subclusters in the electromagnetic (EM) section of the calorimeter
($\pi$-like), 2) single track with EM subclusters ($\rho$-like),
or 3) more than one associated track. No attempt was made to
separate hadrons from electrons (which can contribute to
both $\tau$-type 1 and $\tau$-type 2).
 
Additional requirements (which depend on the $\tau$-type) imposed on the selected events
to enhance the signal-to-background ratio are shown in 
 Table~\ref{tab:tabCuts}.   The background increases rapidly with 
decreasing $p_T^{\mu}$
or decreasing $E_T^{\tau}$. It is significantly lower for $\tau$-type 2 
than for the other $\tau$-types, so a lower $E_T^{\tau}$ cut is warranted for
that $\tau$-type. 
The $|\phi_\mu-\phi_\tau|>2.5$ cut takes advantage of the fact 
that most $Z$ bosons have low $p_T$ and thus the decay $\tau$ leptons are 
back-to-back in $\phi$. 
The longitudinal
shape variable
${\cal R}_{trk}^\tau$ (defined in Table~\ref{tab:tabCuts} caption) 
is used to remove misidentified muons 
because it has
a distribution that peaks at much lower values for muons than for
$\tau$ leptons.

\begin{table}
\caption{\label{tab:tabCuts} Event pre-selection cuts}
\begin{ruledtabular}
\begin{tabular}{lr}
Selection & applied to the $\tau$-types\\
\hline
only one $\mu$ & all \\
$p_T^{\mu}>12$ GeV & all \\
$\mu$ isolation& all \\
$E_T^{\tau}>10(5)$ GeV & 1 and 3 (2)\\
$\Sigma p_T^{\tau_{trk}}>7(5)$ GeV & 1 and 3 (2)\\
$rms_{\tau}<0.25$\footnotemark[1]& all \\
$|\phi_\mu-\phi_\tau|>2.5$& all \\
${\cal R}_{trk}^\tau>0.7$\footnotemark[2] & 1 and 2\\
\end{tabular}
\end{ruledtabular}
\footnotetext[1] {$rms_{\tau}=\sqrt{\sum_{i=1}^{n}[(\Delta\phi_i)^2+(\Delta\eta_i)^2]E_{T_i}/E_T}$, 
where $i=1,..,n$ is the index of the calorimeter tower associated with the $\tau$-cluster; 
$\Delta \eta_i$ and $\Delta \phi_i$ are the $\eta$ and $\phi$ difference
between the center of the $\tau$-cluster and calorimeter tower $i$.}
\footnotetext[2]{${\cal R}_{trk}^\tau=(E^\tau-E_{CH}^{trk})/p_T^{trk}$, where 
$E_{CH}^{trk}$ is the energy deposited in a  window of $5 \times 5$ towers (each
tower of size $\phi \times \eta$=0.1$\times$0.1) around 
the $\tau$-track  in the coarse hadronic (CH) section  of calorimeter.}
\end{table}

The $\tau$ leptons from a $Z$ boson decaying to hadrons + $\nu_{\tau}$
have average visible energy ($E^{\tau}$)
 of the order of 25 GeV and need to be separated from a 
very large background of jets.
To further reduce the jet background, 
a neural network (NN) \cite{NN} consisting of a
single input layer containing several nodes (one for each
input variable),
a single hidden layer with the same number of nodes, and a single output node was used.
A separate NN was trained for each type using a Monte Carlo (MC) sample of
single $\tau$ leptons uniformly distributed in $E_T$ and $\eta$ and overlayed
with a minimum bias event for signal \cite{pythia},
and jets recoiling against non-isolated muons from data for background.
The NN input variables were chosen to minimize the dependence on the $\tau$ energy
and to exploit the narrow width of the energy deposition
in the calorimeter, the low track multiplicity,
the low $\tau$ mass,
and the fact that $\tau$ leptons from $Z$ boson decays are well isolated.
The NN input variables were:

\begin{enumerate}
\item {\it profile} $= (E_{T_1}+E_{T_2})/E_T^{\tau}$, where $E_{T_1}$ 
and $E_{T_2}$ are 
the $E_T$ of the two most energetic calorimeter towers. 
Used for all $\tau$-types. 
\item  {\it caliso} $= (E_T^{\tau}-E_T^{core})/E_T^{core}$. 
A calorimeter isolation parameter used for all $\tau$-types. 
\item $trkiso=\Sigma p_T^{trk}/ \Sigma p_T^{\tau_{trk}}$, where $p_T^{trk}$
($p_T^{\tau_{trk}}$) is
the $p_T$ of a track within a $R<0.5$ 
cone not associated (associated) with the $\tau$ candidate.
A track isolation parameter used for all $\tau$-types.
\item $(E^{EM_1}+E^{EM_2})/E^{\tau}$ in a $R<0.5$ cone, where $E^{EM_1}$ and $E^{EM_2}$ 
are the energies deposited in the first two layers of the EM calorimeter. A parameter 
used for $\tau$-type 1 to reject jets with one energetic charged track and soft $\pi^0$ mesons.
\item $p_T^{\tau_{trk1}}/E_T^{\tau}$, where $p_T^{\tau_{trk1}}$ is $p_T$ 
of the highest $p_T$ track associated with the $\tau$. 
Used for $\tau$-type 1 and 3.
\item $p_T^{\tau_{trk1}}/(E_T^{\tau}\cdot${\it caliso}). A parameter used 
for $\tau$-type 2
that measures the correlation between track and energy deposition 
in isolation annulus.
\item $e_{12}= \sqrt{\Sigma p_T^{\tau_{trk}} \cdot E_T^{EM}}/E_T^{\tau}$, where 
 $E_T^{EM}$ is the transverse energy deposited in the EM layers of the 
calorimeter. Used for $\tau$-types 2 and 3.
\item $\delta \alpha = \sqrt{(\Delta \phi/\sin \theta)^2+(\Delta \eta)^2}$, 
where the differences are between $\Sigma \tau$-tracks and $\Sigma$EM-clusters. 
In the small angle approximation
the observed $\tau$ mass is given by  $e_{12} \cdot E_T^{\tau}\cdot \delta \alpha$.
Used for $\tau$-types 2 and 3.
\end{enumerate} 

The dominant background is from multijet (QCD) processes, mainly 
from $b \bar b$ events  where the muon isolation requirement is met and 
a jet satisfies the $\tau$ selection criteria. The other
sources of background are $W\rightarrow\mu\nu$ + jets
 and $Z/\gamma^*\rightarrow \mu^+ \mu^-$ with
one of the muons misidentified as a $\tau$ lepton.  The 
${\cal R}_{trk}^{\tau}>0.7$ cut removed 70\% of the $\mu^+ \mu^-$ 
background while keeping 98\% of the expected $Z/\gamma^*\rightarrow 
\tau^+ \tau^-$ events. The number of events that did 
not satisfy this criterion was used to estimate the background from misidentified
muons remaining in the sample after the cut. 

The selected 29,021 events were separated into two samples: 
$\mu$ and $\tau$ of 
opposite charge sign (OS), and $\mu$ and $\tau$ of same charge sign (SS). The OS sample
contains the signal. The SS sample is dominated by background and was used
to predict the QCD background distributions
in the signal sample. From detailed
studies of a sample of data with non-isolated muons, we established that this
procedure is sound if one accounts for a small excess of OS over SS
events that varies somewhat with the $\tau$-type. The correction 
factors ($f_i$, where $i$ denotes the $\tau$-type) were determined to 
be $1.06\pm 0.06$, $1.09\pm 0.03$, and $1.03\pm 0.02$, by
taking the ratio of OS to SS data in the non-isolated muon sample.
There was no observable dependence
of $f_i$ as function of $E_T^{\tau}$, of NN output ($NN$) values, or
of the muon parameters. A possible dependence of $f_i$ on the degree
of isolation of muons coming from jets
was checked by looking at the variation in the $f_i$ as function of muon $p_T$
relative to the jet axis and varying the muon isolation.
No variation was observed within the systematic uncertainties
quoted. An overall 3\% systematic uncertainty was added
for the extrapolation to the $NN>0.8$ region.
These factors do not fully account,
however, for the contribution from $W\rightarrow\mu\nu$ + jets, 
which have a larger
excess of OS over SS and different distributions. The additional
contribution of this channel to the signal sample is 
estimated from {\sc pythia} \cite{pythia} MC
samples. The MC
is normalized using the OS and SS data  with $p_T^{\mu}>20$ GeV, 
$|\phi_\mu-\phi_\tau|<2.0$,
and $0.3<NN<0.8$ (in this region $W\rightarrow\mu\nu$ events 
dominate over the  QCD background). The additional contribution to the
 background from $W\rightarrow\tau\nu$ events was ignored
as it is a small fraction of the uncertainty on the contribution
from $W\rightarrow\mu\nu$ events.

\begin{figure}

\includegraphics[scale=0.45]{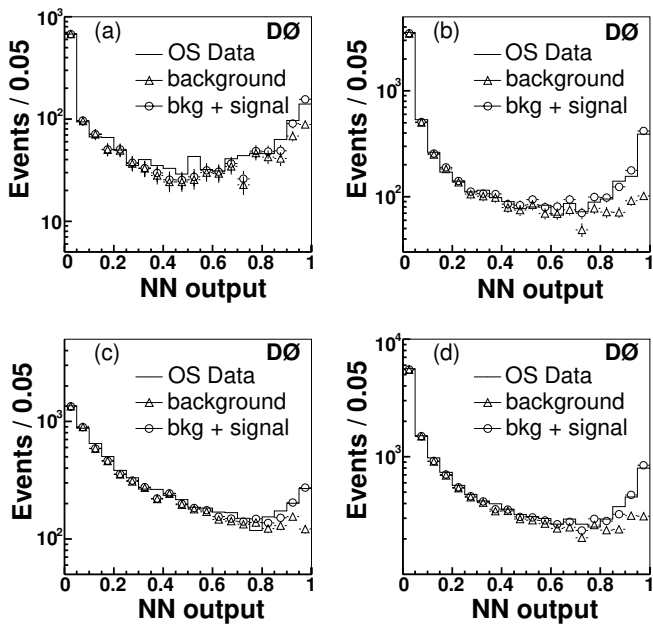}

\caption{\label{fig:NN}NN output distributions for: (a) $\tau$-type 1, (b) $\tau$-type 2,
(c) $\tau$-type 3, and (d) the sum over all the $\tau$-types.}

\end{figure}

\begin{figure}

\includegraphics[scale=0.45]{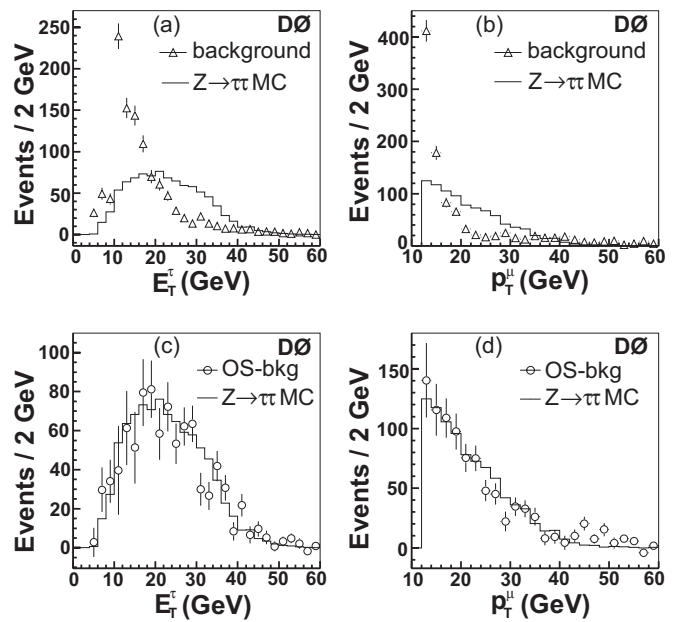}

\caption{\label{fig:ETtau} $E_T^{\tau}$ [(a), (c)] and $p_T^{\mu}$ [(b),(d)] distributions after
$NN>0.8$ cut: (a), (b) estimated background (open triangles)
and predicted $Z\rightarrow \tau \tau$ signal (histogram); 
(c), (d) background subtracted data (open circles)
and predicted $Z\rightarrow \tau \tau$ signal.}

\end{figure}

Figure \ref{fig:NN} shows
the $NN$ distributions for each $\tau$-type (and the sum) for the signal sample, the predicted background and the result of adding the
predicted signal (from $Z/\gamma^*\rightarrow \tau \tau$ MC
\cite{pythia}) to
the background. Table \ref{tab:events} shows the total number
of events observed and predicted before and after the final cut $NN>$0.8.
Distributions of background
subtracted data are in very good agreement with those expected
from $Z\rightarrow \tau \tau$ MC. Figure \ref{fig:ETtau}
compares the expected $E_T^{\tau}$ and  $p_T^{\mu}$ (adding all $\tau$-types)
signal distributions to the predicted background distributions, and to the  
distributions obtained by subtracting the predicted background
from the signal sample distributions.

\begin{table}
\caption{\label{tab:events} Number of predicted and observed contributions  to 
OS events by $\tau$-type before and after the $NN>0.8$ cut}
\begin{ruledtabular}
\begin{tabular}{lcccc}
~ & $\tau$-type 1 & $\tau$-type 2 & $\tau$-type 3 & Total\\
\hline
QCD\footnotemark[1] & $1638\pm 107$ & $6001\pm 187$ & $6242\pm 153$ & $13881\pm 264$ \\
 $Z/\gamma^*\rightarrow\mu\mu$ & $~~33\pm 11~$ & $~~67\pm 22~$ & -- & $~~100\pm 24~$ \\
 $W\rightarrow\mu \nu$\footnotemark[2] & $~~42\pm 41~~$ & $~151\pm 93~$ & $~241\pm 114~$ & $~~434\pm 153$ \\
$Z/\gamma^*\rightarrow \tau \tau$\footnotemark[3] & $~139\pm 6~~$ & $~700\pm 26~$ & $~335\pm 14~$ & $~1174\pm 43~$ \\
\hline
Sum & $1852\pm117$ & $7019\pm 214$ & $6818\pm 189$ & $15589\pm 309$ \\ 
OS events & 1880 & 6971   & 7060 & 15911 \\
\hline
\hline
\multicolumn{5}{c}{$NN>0.8$} \\
\hline
QCD & $~196\pm 23~$ & $~280\pm 24~$ & $~508\pm 32~$ & $~~984\pm 46~$ \\
  $Z/\gamma^*\rightarrow\mu \mu$ & $~~30\pm 10~$ & $~~40\pm 13~$ & -- & $~~~70\pm16~$ \\
 $W\rightarrow\mu \nu$ & $3\pm 5$ & $~~17\pm 11~~$ & $~~38\pm 16~$ & $~~~58\pm 20~$\\
$Z/\gamma^*\rightarrow \tau \tau$ & $~121\pm 6~~$ & $~532\pm 21~$ & $~261\pm 11~$ & $~~914\pm 24~$ \\
\hline
Sum & $~350\pm 26~$ & $~869\pm 36~$ & $~807\pm 37~$ & $~2026\pm 57~$ \\
OS events & 355 & 820 & 833 & 2008 \\
\end{tabular}
\end{ruledtabular}
\footnotetext[1]{The QCD background is estimated by multiplying the
number of SS events by $f_i$ (described in the text).}
\footnotetext[2]{The expected contribution is the number of events that must
be added after subtracting the corrected number of SS events from OS events.}
\footnotetext[3]{The predicted number of $Z/\gamma^*\rightarrow \tau^+ \tau^-$ events 
is based on a theoretical cross section of 257$\pm$9 pb for
$M_{\tau \tau}>60$ GeV \cite{theory} plus 3.5\% predicted from MC for the
number of events expected with $M_{\tau \tau}<60$ GeV.}

\end{table}

\begin{table}
\caption{\label{tab:SysErr} Systematic uncertainties on $\sigma\cdot$Br$(Z/\gamma^*\rightarrow \tau^+ \tau^-)$ }
\begin{ruledtabular}
\begin{tabular}{lr}
Energy Scale & 2.5\% \\
NN & 2.6\% \\
QCD background & 3.5\% \\
  $Z/\gamma^*\rightarrow\mu \mu$ background & 2\% \\ 
$W\rightarrow \mu \nu$ background & 2.3\% \\
$Z/\gamma^*\rightarrow \tau \tau$ MC & 1.5\% \\ 
PDF\footnotemark[1]& 1.7\% \\
$\epsilon_{\text{data}}/\epsilon_{MC}$\footnotemark[2] & 2.1\% \\
Trigger & 3.5\% \\
\hline
Total & 7.5\% \\
\end{tabular}
\end{ruledtabular}
\footnotetext[1]{Efficiency uncertainty due to uncertainty in parton distribution
function (PDF).}
\footnotetext[2]{ $\epsilon_{\text data}/\epsilon_{MC}$ is the ratio of data to MC reconstruction
efficiency.}
\end{table}

The total event efficiency ($\epsilon_{\rm TOT}$) summed over $\tau$-types 1, 2, and 3 is
$1.52\%$ for $M_{\tau \tau}$ greater than
60 GeV . The total efficiency accounts for all losses due to
branching ratios, geometrical acceptance, reconstruction and trigger
efficiencies. It is corrected for the small difference
between MC and data reconstruction efficiencies.
The contributions of the three $\tau$-types to the signal in the final
data sample are $13\%$, $58\%$, and  $29\%$.  

The cross section times branching ratio for $Z/\gamma^*\rightarrow \tau^+ \tau^-$ is given by
$N_{\text{signal}}$/($\epsilon_{\rm TOT}\cdot \int{\cal L}dt$) where $N_{\text{signal}}$ is
given by the number of signal events and
$\int{\cal L}dt$ is the integrated luminosity 
of the sample studied.
$N_{\text{signal}}=865\pm$55 (statistical uncertainty only) is the number of OS events 
of all $\tau$ types after selecting the events with $NN>0.8$, subtracting 
the estimated background (see Table \ref{tab:events}), and subtracting
the number of expected events in the sample with $M_{\tau \tau}$ less than
60 GeV (3.5\%).
  
\begin{figure}

\includegraphics[scale=0.45]{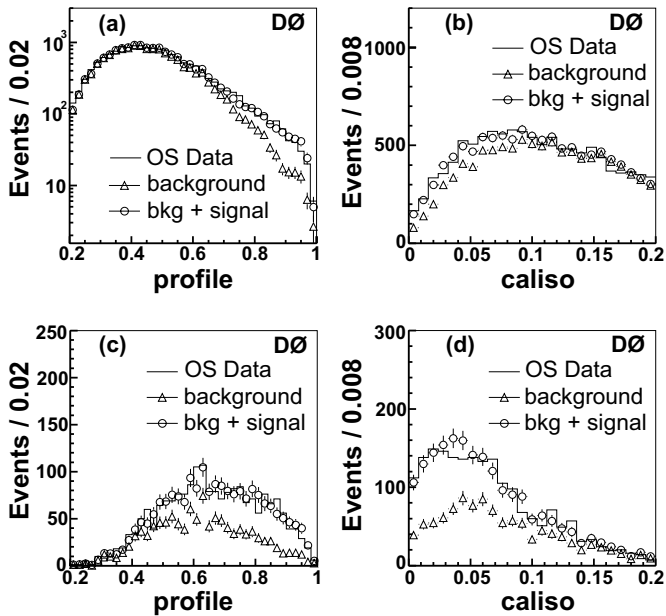}

\caption{\label{fig:NNvar} Distributions for OS data, background and background plus
signal of two NN input variables before  [(a), (b)] and after  
[(c), (d)]$NN>0.8$ cut: (a),(c) {\it profile}; (b),(d) {\it caliso}.}
\end{figure}

The systematic uncertainties on the cross section measurement 
are listed in Table \ref{tab:SysErr}.
The uncertainty (2.5\%) due to the energy scale was estimated 
from the change in the acceptance
when scaling the energy in MC events by the energy
difference between MC and data (as determined by the $p_T$ imbalance
in photon + jet events).
The systematic uncertainty due to the NN performance (2.6\%) 
was estimated by generating ensembles of Monte
Carlo events in which the number of events in each bin of
distributions of NN input variables was allowed
to fluctuate by the uncertainties in the difference between
MC distributions and the background-subtracted data distributions.
The distributions of NN input variables are in good
agreement with those predicted adding $Z/\gamma^*\rightarrow \tau^+ \tau^-$ MC
and the estimated background;
two are shown in Fig. \ref{fig:NNvar}. 

The QCD systematic uncertainty (3.5\%) is due to the uncertainty in determining $f_i$.
The uncertainty in the $Z/\gamma^*\rightarrow\mu^+\mu^-$ and  $W\rightarrow \mu \nu$ 
backgrounds (2.0\% and 2.3\%) come from the statistical uncertainty in determining their
contribution, while the $Z/\gamma^*\rightarrow \tau^+ \tau^-$ MC
systematic uncertainty reflects limited signal MC statistics. 
The $\epsilon_{\text{data}}/\epsilon_{MC}$ is dominated by the uncertainty
in estimating the difference in $\tau$-type 3 tracking efficiency between MC and data
using the the ratio of two- to three-prong events  between background 
subtracted data and
$Z/\gamma^*\rightarrow \tau^+ \tau^-$ MC in
$\tau$-type 3 candidates. The uncertainty from differences in subcluster
reconstruction, single isolated track reconstruction and muon isolation
add up to about 1\%.

The trigger efficiencies were estimated using $Z\rightarrow \mu^+\mu^-$ data,
the systematic uncertainty comes from the statistical uncertainty in that data;
the uncertainties include dependencies on $\eta$ and $\phi$.
Systematic uncertainties from all other sources are less than 1\%.
Thus we obtain
$$\sigma \cdot {\rm Br}(Z/\gamma^*\rightarrow \tau \tau)=252\pm16(\rm{stat})
\pm19(\rm{sys}){\rm ~pb}$$
for $M_{\tau \tau}$ greater than 60 GeV.
The quoted statistical uncertainty is the uncertainty from OS and SS statistics (excluding the uncertainties on 
the correction factors). This yields, after removing the $\gamma^*$ contribution,
$$\sigma \cdot {\rm Br}(Z\rightarrow \tau \tau)=237\pm15(\rm{stat})\pm18(\rm{sys})
\pm15(\rm{lum}){\rm ~pb} $$
in good agreement with the NNLO standard model prediction of 242$\pm$9 pb 
\cite{theory}.

%
We thank the staffs at Fermilab and collaborating institutions, 
and acknowledge support from the 
Department of Energy and National Science Foundation (USA),  
Commissariat  \` a l'Energie Atomique and 
CNRS/Institut National de Physique Nucl\'eaire et 
de Physique des Particules (France), 
Ministry of Education and Science, Agency for Atomic 
   Energy and RF President Grants Program (Russia),
CAPES, CNPq, FAPERJ, FAPESP and FUNDUNESP (Brazil),
Departments of Atomic Energy and Science and Technology (India),
Colciencias (Colombia),
CONACyT (Mexico),
KRF (Korea),
CONICET and UBACyT (Argentina),
The Foundation for Fundamental Research on Matter (The Netherlands),
PPARC (United Kingdom),
Ministry of Education (Czech Republic),
Canada Research Chairs Program, CFI,
Natural Sciences and Engineering Research Council and 
WestGrid Project (Canada),
BMBF and DFG (Germany),
A.P.~Sloan Foundation,
Research Corporation,
Texas Advanced Research Program,
and the Alexander von Humboldt Foundation.
%

\end{document}